\documentclass{sciposter}
\usepackage{lipsum}
\usepackage{epsfig}
\usepackage{amsmath}
\usepackage{amssymb}
\usepackage{multicol}
\usepackage{graphicx,url}
\usepackage[brazil]{babel}   
\usepackage[utf8]{inputenc}
\usepackage{hyperref}
\usepackage{caption}

\hypersetup{
    colorlinks=true,       % false: boxed links; true: colored links
    linkcolor=red,          % color of internal links (change box color with linkbordercolor)
    citecolor=green,        % color of links to bibliography
    filecolor=magenta,      % color of file links
    urlcolor=blue           % color of external links
}

\addto\captionsbrazil{}

\title{Constru\c{c}\~ao e Publica\c{c}\~ao de Artigos Cient\'ificos em Computa\c{c}\~ao}

\author{\href{http://www.wladmirbrandao.com}{Wladmir C. Brand\~ao}}

\institute 
{Departamento de Ci\^encia da Computa\c{c}\~ao \\ Pontif\'icia Universidade Cat\'olica de Minas Gerais (PUC Minas)
}

\email{\href{mailto:wladmir@pucminas.br}{wladmir@pucminas.br}}

\rightlogo[1]{}
\leftlogo[1]{}
\begin{document}

% \begin{textblock*}{100px}(0.98\textwidth,2pt)
%   \includegraphics[width=1.0\textwidth]{QRCode.png}
% \end{textblock*}

\conference{\bf \href{http://www.wladmirbrandao.com}{www.wladmirbrandao.com}}

%\LEFTSIDEfootlogo  
% Uncomment to put footer logo on left side, and 
% conference name on right side of footer

% Some examples of caption control (remove % to check result)

%\renewcommand{\algorithmname}{Algoritme} % for Dutch

%\renewcommand{\mastercapstartstyle}[1]{\textit{\textbf{#1}}}
%\renewcommand{\algcapstartstyle}[1]{\textsc{\textbf{#1}}}
%\renewcommand{\algcapbodystyle}{\bfseries}
%\renewcommand{\thealgorithm}{\Roman{algorithm}}

\maketitle

%%% Begin of Multicols-Enviroment
\begin{multicols}{3}

%%% Abstract
\begin{abstract}
%-- Problema
Ao longo de 15 anos de doc\^encia orientando discentes, participando de bancas de conclus\~ao de curso e coordenando projetos de pesquisa e atividades de inicia\c{c}\~ao cient\'ifica em computa\c{c}\~ao, tenho observado as dificuldades dos discentes na constru\c{c}\~ao de artigos cient\'ificos para apresenta\c{c}\~ao de seus resultados de pr\'aticas de pesquisa, al\'em dos recorrentes questionamentos sobre o processo de publica\c{c}\~ao.
%-- Solucao
Diante disso, o presente artigo procura clarear a quest\~ao, propondo  um arcabou\c{c}o conceitual para suportar a constru\c{c}\~ao e publica\c{c}\~ao de artigos cient\'ificos em computa\c{c}\~ao.
Assim, os discentes da \'area ter\~ao \`a sua disposi\c{c}\~ao uma esp\'ecie de guia para apresenta\c{c}\~ao mais efetiva de resultados de suas pr\'aticas de pesquisa, sobretudo as de car\'ater experimental.

\end{abstract}

\section{Artigo Cient\'ifico em Computa\c{c}\~ao}
\label{sec:artigo}
%-- O que e um artigo cientifico?
Um artigo cient\'ifico \'e o instrumento utilizado por cientistas e pesquisadores para compartilhamento e dissemina\c{c}\~ao do conhecimento cient\'ifico.
Na computa\c{c}\~ao, o artigo cient\'ifico \'e frequentemente concebido como um documento, geralmente composto por texto e imagens, com a finalidade de apresentar resultados experimentais de pr\'aticas de pesquisa para um p\'ublico qualificado, majoritariamente composto por cientistas, pesquisadores, profissionais, professores e estudantes da \'area.

%-- Quais os aspectos importantes em um artigo cientifico?
Uma estrat\'egia comumente utilizada por cientistas e pesquisadores de computa\c{c}\~ao para apresenta\c{c}\~ao de resultados experimentais, provenientes de suas pr\'aticas de pesquisa, \'e a ado\c{c}\~ao de narrativa simples e direta que \textit{``conta uma hist\'oria''} sobre um tema, ao mesmo tempo em que \textit{``vende uma ideia''} para solucionar um problema.
O objetivo \'e atrair a aten\c{c}\~ao e convencer o p\'ublico da relev\^ancia da ideia para a solu\c{c}\~ao do problema computacional.

Para demonstrar a relev\^ancia da ideia, \'e imprescind\'ivel a presen\c{c}a de aspectos como originalidade, impacto, objetividade e organiza\c{c}\~ao na narrativa do artigo.
A seguir s\~ao apresentadas algumas perguntas, divididas por aspectos, que devem ser utilizadas por autores de artigos cient\'ificos em computa\c{c}\~ao para qualificar a relev\^ancia de suas ideias:

\begin{itemize}
  \item[$\triangleright$] \textsc{Originalidade}: O problema a ser solucionado pela ideia \'e \'unico? A ideia para a solu\c{c}\~ao do problema \'e criativa e inovadora?
  \item[$\triangleright$] \textsc{Impacto}: A ideia implicar\'a desdobramento em m\'ultiplas \'areas de pesquisa? A ideia pode influenciar significativamente a forma de pensar e agir do p\'ublico e da sociedade em geral? O artigo tem potencial para gerar cita\c{c}\~oes?
  \item[$\triangleright$] \textsc{Objetividade}: As ideias s\~ao apresentadas de maneira clara, objetiva e completa? O discurso \'e composto exclusivamente por argumentos imprescind\'iveis para o entendimento das ideias e de seu impacto?
    \begin{itemize}
      \item[$\triangleright$] \textsc{Linguagem}: H\'a uso adequado da linguagem?
      \item[$\triangleright$] \textsc{Apontamentos}: H\'a preocupa\c{c}\~ao em citar outros artigos cient\'ificos necess\'arios e suficientes para a compreens\~ao das ideias e estabelecimento do estados-da-arte? Os apontadores est\~ao explicitamente indicados no texto de maneira adequada?
      \item[$\triangleright$] \textsc{Reprodutibilidade}: A descri\c{c}\~ao da ideia est\'a detalhada e com precis\~ao suficiente para que outros pesquisadores consigam reproduzir a pesquisa? 
      \item[$\triangleright$] \textsc{Solidez}: Os experimentos foram planejados e executados adequadamente? As an\'alises sobre os resultados s\~ao profundas e esclarecedoras? Testes estat\'isticos para valida\c{c}\~ao experimental dos resultados foram efetuados? 
    \end{itemize}    
  \item[$\triangleright$] \textsc{Organiza\c{c}\~ao}: A ideia \'e apresentada com encadeamento l\'ogico intelig\'ivel ao p\'ublico? A \textit{``hist\'oria contada''} apresenta um bom \textit{``enredo''} para \textit{``vender a ideia''}?
\end{itemize}

Os questionamentos apresentados acima devem funcionar como instrumento de avalia\c{c}\~ao do potencial do artigo cient\'ifico atrair a aten\c{c}\~ao e se mostrar relevante para o p\'ublico. 
Al\'em disso, os autores devem ter em mente que a capta\c{c}\~ao desses quatro aspectos, que influenciam diretamente na qualidade e no potencial de impacto de um artigo cient\'ifico, \'e feita a partir de tr\^es dimens\~oes distintas presentes no documento: conte\'udo, forma e apontadores.

\columnbreak
\section{Conte\'udo do Artigo Cient\'ifico}
\label{sec:conteudo}
A dimens\~ao \textit{conte\'udo} do artigo cient\'ifico lida com a descri\c{c}\~ao do tema de pesquisa e da ideia apresentada para a solu\c{c}\~ao do problema. Essa dimens\~ao pode ser desmembrada em nove elementos:

\begin{itemize}
  \item[$\triangleright$] \textsc{Motiva\c{c}\~ao}: Qual a relev\^ancia do tema de pesquisa para a sociedade?
  \item[$\triangleright$] \textsc{Problema}: Qual a demanda a ser atendida pela aplica\c{c}\~ao da ideia? Que problema de pesquisa a ideia soluciona? Qual o conjunto de hip\'oteses da pesquisa?
  \item[$\triangleright$] \textsc{Justificativa}: Qual a import\^ancia de se solucionar o problema? Qual o impacto e quais os desdobramentos em pesquisa podem surgir como fruto da aplica\c{c}\~ao da ideia para a solu\c{c}\~ao do problema?
  \item[$\triangleright$] \textsc{Objetivo}: Quais s\~ao as proposi\c{c}\~oes e a\c{c}\~oes executadas para se solucionar o problema?
  \item[$\triangleright$] \textsc{Solu\c{c}\~ao}: Como solucionar o problema? Qual a especifica\c{c}\~ao da ideia para se solucionar o problema?
  \item[$\triangleright$] \textsc{Comprova\c{c}\~ao}: Qual a capacidade pr\'atica da ideia e qu\~ao efetiva ela \'e para solucionar o problema?
  \item[$\triangleright$] \textsc{Contribui\c{c}\~ao}: Quais os resultados alcan\c{c}ados? Qu\~ao original e impactante \'e a ideia? Quais s\~ao os produtos resultantes da comprova\c{c}\~ao e da aplica\c{c}\~ao da ideia para solucionar o problema?
  \item[$\triangleright$] \textsc{Conclus\~ao}: Qual \'e a s\'intese do problema e da solu\c{c}\~ao? Quais hip\'oteses foram confirmadas e quais foram refutadas?
  \item[$\triangleright$] \textsc{Dire\c{c}\~oes Futuras}: Que novos problemas e ideias para sua solu\c{c}\~oes podem ser apontados a partir da pesquisa?
\end{itemize}

\section{Forma do Artigo Cient\'ifico}
\label{sec:forma}
A dimens\~ao \textit{forma} lida com a segmenta\c{c}\~ao e disposi\c{c}\~ao do conte\'udo em uma estrutura apropriada, privilegiando o aspecto de organiza\c{c}\~ao e o encadeamento l\'ogico da argumenta\c{c}\~ao sobre a ideia. Em geral, essa estrutura apresenta os seguintes elementos:

\begin{itemize}
  \item[$\triangleright$] \textsc{T\'itulo}: Conjunto de termos usados para designar o artigo cient\'ifico. Frequentemente, uma frase que sintetiza o objetivo \'e usada como t\'itulo. Tamb\'em \'e comum \textit{``batizar''} a solu\c{c}\~ao proposta com um acr\^onimo que remeta ao objetivo, e utilizar o acr\^onimo e seu respectivo significado como t\'itulo do artigo.
  \item[$\triangleright$] \textsc{Autores}: Lista de nomes de autores e suas respectivas afilia\c{c}\~oes.
  \item[$\triangleright$] \textsc{Resumo}: Descri\c{c}\~ao sucinta do conte\'udo, enfatizando o problema, o objetivo, a solu\c{c}\~ao e a contribui\c{c}\~ao.
  \item[$\triangleright$] \textsc{Introdu\c{c}\~ao}: Descri\c{c}\~ao ampliada do conte\'udo, apresentando elementos de motiva\c{c}\~ao, problema, justificativa, objetivo, contribui\c{c}\~ao, solu\c{c}\~ao e comprova\c{c}\~ao. Al\'em disso, apresenta descri\c{c}\~ao sucinta da sequ\^encia de se\c{c}\~oes de conte\'udo que ser\~ao encontradas ao longo do artigo.
  \item[$\triangleright$] \textsc{Referencial Te\'orico}: Apresenta\c{c}\~ao de conceitos referenciados na literatura cient\'ifica relacionados ao tema que ajudam o p\'ublico a entender o problema e as ideias contidas no artigo.
  \item[$\triangleright$] \textsc{Trabalhos Relacionados}: Apresenta\c{c}\~ao de ideias publicadas na literatura cient\'ifica para solu\c{c}\~ao do mesmo problema tratado no artigo (estado-da-arte), a fim de estabelecer uma linha de base comparativa para a solu\c{c}\~ao apresentada.
  \item[$\triangleright$] \textsc{Metodologia}: Descrição detalhada da solu\c{c}\~ao apresentada para o problema, incluindo modelos, algoritmos, abordagens, m\'etodos e t\'ecnicas.
  \item[$\triangleright$] \textsc{Experimentos e Resultados}: Descri\c{c}\~ao da comprova\c{c}\~ao envolvendo ambiente experimental, bases de dados artificiais, instrumentos e mecanismos de coleta de dados, ferramentas para tratamento de dados, m\'etricas utilizadas em compara\c{c}\~oes, valida\c{c}\~oes estat\'isticas e an\'alises sobre os resultados.
  \item[$\triangleright$] \textsc{Conclus\~ao e Trabalhos Futuros}: Descri\c{c}\~ao resumida do problema, dos objetivos, da solu\c{c}\~ao, da comprova\c{c}\~ao, da contribui\c{c}\~ao e das dire\c{c}\~oes futuras.
  \item[$\triangleright$] \textsc{Referencial Bibliogr\'afico}: Lista de refer\^encias para artigos publicados na literatura cient\'ifica que foram apontadas em outros elementos da estrutura do artigo.
\end{itemize}

\section{Apontadores do Artigo Cient\'ifico}
\label{sec:apontamentos}
A dimens\~ao \textit{apontadores} lida com a especifica\c{c}\~ao e utiliza\c{c}\~ao de refer\^encias bibliogr\'aficas e cita\c{c}\~oes a essas refer\^encias.
No artigo cient\'ifico, a refer\^encia \`as ideias previamente publicadas por outros cientistas e pesquisadores deve sempre vir acompanhada de um apontador (cita\c{c}\~ao) para o registro bibliogr\'afico do artigo publicado (refer\^encia bibliogr\'afica).
%-- Credibilidade 
\'E imprescind\'ivel a correta utiliza\c{c}\~ao de apontadores que indiquem precisamente a origem da ideia referenciada. 
Al\'em disso, a adequa\c{c}\~ao das refer\^encias bibliogr\'aficas \`as normas preestabelecidas pela comunidade cient\'ifica influencia positivamente a organiza\c{c}\~ao do artigo.

O desafio para todo autor de artigo cient\'ifico consiste na constru\c{c}\~ao de um referencial bibliogr\'afico adequado que forne\c{c}a suporte \`a exposi\c{c}\~ao objetiva de suas ideias. 
Diante desse desafio, surgem algumas quest\~oes: 
\begin{enumerate}
  \item Onde encontrar artigos cient\'ificos para a constru\c{c}\~ao do referencial bibliogr\'afico?
  \item Como filtrar os artigos cient\'ificos relevantes para a constru\c{c}\~ao do referencial bibliogr\'afico?
  \item Como identificar dentre os artigos cient\'ificos encontrados aqueles mais adequados para cita\c{c}\~ao?
  \item Como especificar o referencial bibliogr\'afico e realizar cita\c{c}\~oes no artigo cient\'ifico?
\end{enumerate}

Nas se\c{c}\~oes subsequentes abordaremos cada uma das quest\~oes citadas previamente. A Se\c{c}\~ao~\ref{sec:apontamentos-fontes} aborda a quest\~ao 1, a Se\c{c}\~ao~\ref{sec:apontamentos-aquisicao} aborda a quest\~ao 2, a Se\c{c}\~ao~\ref{sec:apontamentos-qualidade} aborda a quest\~ao 3 e a Se\c{c}\~ao~\ref{sec:apontamentos-normalizacao} aborda a quest\~ao 4.\\

\subsection{Fontes de Informa\c{c}\~ao Bibliog\'afica}
\label{sec:apontamentos-fontes}
Existem in\'umeras fontes de informa\c{c}\~ao bibliogr\'afica que podem ser utilizadas para encontrar documentos relevantes para a constru\c{c}\~ao de referencial bibliogr\'afico.
Especificamente em computa\c{c}\~ao, algumas dessas fontes est\~ao dispon\'iveis para consulta \textit{online} e s\~ao amplamente utilizadas pela comunidade cient\'ifica. Dentre elas, destacam-se:

\begin{itemize}
  \item[$\triangleright$] \textsc{\href{http://www.dl.acm.org}{ACM Digital Library}}: Biblioteca digital da \textit{\href{http://www.acm.org}{Association for Computing Machinery (ACM)}}, a maior comunidade cient\'ifica e educacional em computa\c{c}\~ao do mundo.
  \item[$\triangleright$] \textsc{\href{http://ieeexplore.ieee.org}{IEEE Xplore}}: Biblioteca digital da \textit{\href{http://www.ieee.org}{Institute of Electrical and Electronics Engineers (IEEE)}}, a maior associa\c{c}\~ao profissional mundial constitu\'ida por engenheiros, cientistas e profissionais de computa\c{c}\~ao, eletr\^onica e telecomunica\c{c}\~oes.
  \item[$\triangleright$] \textsc{\href{http://dblp.uni-trier.de}{DBLP}}: A \textit{Digital Bibliography \& Library Project (DBLP)} \'e um \'indice bibliogr\'afico para a ci\^encia da computa\c{c}\~ao desenvolvido e provido pela \textit{\href{http://www.uni-trier.de}{Universit\"at Trier}}.
  \item[$\triangleright$] \textsc{\href{http://citeseerx.ist.psu.edu}{CiteSeer}}: Biblioteca digital mantida pela \textit{\href{http://www.nsf.gov}{National Science Foundation (NSF)}} contendo uma base indexada de artigos em ci\^encia da computa\c{c}\~ao e ci\^encia da informa\c{c}\~ao.
  \item[$\triangleright$] \textsc{\href{http://webofknowledge.com}{Web of Science}}: Servi\c{c}o de indexa\c{c}\~ao de cita\c{c}\~oes cient\'ificas da \textit{\href{http://thomsonreuters.com}{Thomson Reuters}}, contendo uma base indexada de artigos de diferentes campos cient\'ificos.
  \item[$\triangleright$] \textsc{\href{http://www.scopus.com}{Scopus}}: Base de dados bibliogr\'afica mantida pela \textit{\href{http://www.elsevier.com}{Elsevier}} contendo \'indices bibliogr\'aficos de diferentes campos cient\'ificos.
  \item[$\triangleright$] \textsc{\href{http://scholar.google.com}{Google Scholar}}: M\'aquina de busca da \textit{\href{http://google.com}{Google}} para recupera\c{c}\~ao de produ\c{c}\~oes bibliogr\'aficas em diferentes campos cient\'ificos.
\end{itemize}

\subsection{Aquisi\c{c}\~ao de Informa\c{c}\~ao Bibliog\'afica}
\label{sec:apontamentos-aquisicao}
O processo de aquisi\c{c}\~ao de informa\c{c}\~ao bibliogr\'afica consiste, em grande parte, na execu\c{c}\~ao de atividades de busca e filtragem. 
No presente artigo, o processo de aquisi\c{c}\~ao de informa\c{c}\~ao bibliogr\'afica n\~ao ser\'a exaustivamente discutido, mas ser\'a apresentada, de maneira sucinta, a forma como as atividades de busca e filtragem s\~ao comumente executadas nas fontes de informa\c{c}\~ao \textit{online} citadas anteriormente.\\ 

\textsc{Busca}: Existem m\'ultiplas formas de se buscar informa\c{c}\~ao bibliogr\'afica. 
Em geral, a busca em fontes de informa\c{c}\~ao \textit{online} \'e realizada a partir da submiss\~ao de consultas ao mecanismo de busca provido pela fonte para posterior tratamento dos resultados obtidos. Com frequ\^encia, as consultas s\~ao formuladas utilizando palavras-chave relacionadas ao tema de pesquisa (consultas por t\'opico), aos nomes de autores de refer\^encia no tema de pesquisa (consultas por autor) e aos ve\'iculos onde artigos relacionados ao tema s\~ao comumente publicados (consultas por ve\'iculo).\\

\textsc{Filtragem}: A atividade de filtragem \'e usualmente executada atrav\'es da inspec\~ao visual do conte\'udo dos artigos encontrados na busca, a fim de selecionar aqueles considerados relevantes. Devido ao alto custo de inspe\c{c}\~ao visual do texto completo de uma grande quantidade de documentos, frequentemente a filtragem \'e executada em duas fases. Numa primeira fase de inspe\c{c}\~ao r\'apida, os elementos t\'itulo, resumo e palavras-chave dos documentos encontrados na busca s\~ao inspecionados para se decidir se o documento \'e potencialmente relevante para ser selecionado para a pr\'oxima fase. Na fase de inspe\c{c}\~ao profunda, o texto completo dos documentos selecionados na inspe\c{c}\~ao r\'apida \'e analisado a fim de se descartar os pouco relevantes ou irrelevantes. \\

De fato, a filtragem se baseia fortemente na estimativa da qualidade da informa\c{c}\~ao bibliogr\'afica. 
Sendo assim, diferentes crit\'erios al\'em da an\'alise de conte\'udo podem ser adotados para tal fim. 
Na pr\'oxima se\c{c}\~ao, aspectos relacionados a essa estimativa ser\~ao apresentados.\\

\subsection{Qualidade da Informa\c{c}\~ao Bibliog\'afica}
\label{sec:apontamentos-qualidade}
A estimativa de qualidade da informa\c{c}\~ao bibliogr\'afica depende fundamentalmente de um julgamento de valor, o qual deve ser realizado a partir de crit\'erios objetivos.
Em particular, tanto crit\'erios intratextuais, relacionados ao conte\'udo do artigo, como extratextuais, relacionados \`a reputa\c{c}\~ao dos autores e dos ve\'iculos de publica\c{c}\~ao, podem ser utilizados.
Al\'em disso, dada a natureza din\^amica da computa\c{c}\~ao, onde uma parte significativa das ideias publicadas se torna rapidamente obsoleta, a rec\^encia do artigo cient\'ifico tamb\'em se torna um crit\'erio importante para a estimativa.

Para a an\'alise de conte\'udo, os questionamentos apresentados na Se\c{c}\~ao~\ref{sec:conteudo} podem ser utilizados para o estabelecimento dos elementos de an\'alise. 
Para a an\'alise de componentes extratextuais, existem diferentes \'indices e m\'etricas propostos na literatura que podem ser utilizados para o estabelecimento do grau de reputa\c{c}\~ao de autores e de ve\'iculos de publica\c{c}\~ao, muitos deles baseados em n\'umeros de cita\c{c}\~oes.
Existem algumas m\'etricas comumente utilizadas pela comunidade cient\'ifica:

\begin{itemize}
  \item[$\triangleright$] \textsc{N\'umero de cita\c{c}\~oes}: N\'umero de artigos publicados, por autor ou ve\'iculo, citados em outros artigos publicados. 
  \item[$\triangleright$] \textsc{\'Indice h}: Maior n\'umero $h$, sendo que $h$ artigos publicados, por autor ou ve\'iculo, tenha ao menos $h$ cita\c{c}\~oes. Por exemplo, um ve\'iculo de publica\c{c}\~ao com cinco artigos citados por $20$, $8$, $5$, $3$ e $1$ outros artigos, respectivamente, tem um \textit{\'indice h} de $3$. Da mesma forma, um autor que tenha publicado cinco artigos citados por $20$, $8$, $5$, $3$ e $1$ outros artigos, respectivamente, tem um \textit{\'indice h} de $3$.
  \item[$\triangleright$] \textsc{\'Indice i10}: N\'umero de artigos publicados, por autor ou ve\'iculo, com pelo menos 10 cita\c{c}\~oes. 
\end{itemize}

Existem mecanismos que disponibilizam livremente esses \'indices e m\'etricas na Web, tal como o \href{http://scholar.google.com}{Google Scholar}.
Al\'em disso, para estimativa da reputa\c{c}\~ao de ve\'iculos de publica\c{c}\~ao, pode-se utilizar a taxonomia \href{http://qualis.capes.gov.br}{Qualis} disponibilizada pelo  governo brasileiro, por meio da \href{http://www.capes.gov.br}{Coordena\c{c}\~ao de Aperfei\c{c}oamento de Pessoal de N\'ivel Superior (CAPES)}.
A CAPES periodicamente utiliza um conjunto de procedimentos para estratifica\c{c}\~ao da qualidade da produ\c{c}\~ao intelectual dos programas brasileiros de p\'os-gradua\c{c}\~ao, fornecendo uma taxonomia para peri\'odicos, confer\^encias, simp\'osios e workshops nacionais e internacionais na \'area de Ci\^encia da Computa\c{c}\~ao (Qualis-CC) composta de sete categorias: $A1$, $A2$, $B1$, $B2$, $B3$, $B4$ e $C$. 
Pode-se considerar que ve\'iculos classificados no extrato $A1$ tem reputa\c{c}\~ao mais elevada que ve\'iculos classificados no extrato $A2$, que por sua vez tem reputa\c{c}\~ao mais elevada que ve\'iculos classificados no extrato $B1$, e assim sucessivamente.\\

\subsection{Normaliza\c{c}\~ao Bibliogr\'afica}
\label{sec:apontamentos-normalizacao}
Normaliza\c{c}\~ao bibliogr\'afica \'e o processo de adequa\c{c}\~ao das refer\^encias bibliogr\'aficas e cita\c{c}\~oes em respeito \`as normas preestabelecidas, tais como \href{http://www.acm.org/publications/article-templates/acm-latex-style-guide}{ACM}, \href{http://www.ieee.org/documents/ieeecitationref.pdf}{IEEE} e \href{http://www.abnt.org.br}{ABNT}.
Em geral, cada ve\'iculo de publica\c{c}\~ao estabelece quais normas devem ser seguidas para normaliza\c{c}\~ao bibliogr\'afica dos artigos cient\'ificos a serem publicados por eles.
Al\'em disso, frequentemente os ve\'iculos oferecem modelos que podem ser utilizados para facilitar o processo de normaliza\c{c}\~ao.
Em particular, recomenda-se a utiliza\c{c}\~ao de \href{http://www.latex-project.org}{\LaTeX}~para a edi\c{c}\~ao do artigo cient\'ifico, uma vez que grande parte dos ve\'iculos de publica\c{c}\~ao na \'area de computa\c{c}\~ao oferece modelos para essa plataforma.
Mais ainda, a utiliza\c{c}\~ao dessa plataforma facilita a adequa\c{c}\~ao das refer\^encias bibliogr\'aficas e cita\c{c}\~oes \`as diferentes normas.

\section{Recomenda\c{c}\~oes para Constru\c{c}\~ao do Artigo}
\label{sec:dicas}
A seguir s\~ao apresentadas algumas recomenda\d{d}\~oes importantes para a constru\c{c}\~ao de artigos cient\'ificos em computa\c{c}\~ao:

\begin{itemize}
  \item[$\triangleright$] Use \href{http://www.latex-project.org}{\LaTeX}.
  \item[$\triangleright$] Existem m\'ultiplas formas de se especificar e descrever ideias. Particularmente na computa\c{c}\~ao, ideias são comumente especificadas como algoritmos, t\'ecnicas, abordagens, m\'etodos, arquiteturas ou modelos.
  \item[$\triangleright$] Dedique aten\c{c}\~ao especial ao t\'itulo e ao resumo. Frequentemente eles s\~ao repons\'aveis por despertar o interesse do p\'ublico pelo artigo cient\'ifico, uma vez que esse p\'ublico decide se vai ou n\~ao consumir as ideias ali presentes a partir desses elementos.
\item[$\triangleright$] Um aspecto importante \'e a normaliza\c{c}\~ao, que descreve as regras para formata\c{c}\~ao dos elementos estruturais. A obedi\^encia a essas regras muitas vezes \'e imperativa para a aceita\c{c}\~ao do artigo cient\'ifico por ve\'iculos para publica\c{c}\~ao.
  \item[$\triangleright$] Trabalhos relacionados devem ser os mais recentes poss\'iveis, contemplando o estado da arte sobre o tema tratado pelo artigo.
  \item[$\triangleright$] No referencial te\'orico e nos trabalhos relacionados, a qualidade das refer\^encias utilizadas influi na qualidade atribu\'ida ao artigo cient\'ifico pelo p\'ublico. A reputa\c{c}\~ao da fonte de informa\c{c}\~ao reflete na forma como o p\'ublico atribuir\'a qualidade ao artigo cient\'fico. \'E o efeito \textit{``diga-me com quem andas e direi quem \'es''}.
  \item[$\triangleright$] Afirma\c{c}\~oes fortes e categ\'oricas devem apresentar apontadores para as refer\^encias bibliogr\'aficas que as comprovem.
  \item[$\triangleright$] Se\c{c}\~oes muito curtas, com um ou dois par\'grafos, devem ser evitadas.
\end{itemize}

\section{Processo de Publica\c{c}\~ao do Artigo Cient\'ifico}
\label{sec:publicacao}
A charge de Mischa Richter, publicada no \textit{The New Yorker} em 28 de maio de 1966 e apresentada na Figura~\ref{fig:publishperish}, refere-se a um ditado muito popular no meio acad\^emico: publique ou pere\c{c}a.
A met\'afora ali presente remete a um fato conhecido entre pesquisadores: uma \'otima ideia n\~ao divulgada \'e um p\'essimo neg\'ocio para seus autores e para a sociedade.
Da\'i, a relev\^ancia de autores e pesquisadores se empenharem para a publica\c{c}\~ao de seus artigos. 
Por isso, vamos descrever nessa se\c{c}\~ao, ainda que de forma sucinta, o processo convencional de publica\c{c}\~ao de artigos cient\'ificos em computa\c{c}\~ao. 
Esse processo envolve sete atividades distintas, conforme pode ser observado na Figura~\ref{fig:processopublicacao}.
A primeira atividade \'e a sele\c{c}\~ao de ve\'iculos-alvo apropriados para a submiss\~ao do artigo.
Considerando que cada ve\'iculo estabelece crit\'erios pr\'oprios para aceita\c{c}\~ao de artigos, os autores devem estar atentos a alguns aspectos importantes comumente considerados por todos eles:

\begin{itemize}
  \item[$\triangleright$] \textsc{Tema}: Os ve\'iculos restringem o conte\'udo aceito para publica\c{c}\~ao a um conjunto de temas espec\'ificos. Os autores devem se certificar de que o tema tratado pelo artigo seja aderente aos temas de interesse do ve\'iculo.
  \item[$\triangleright$] \textsc{Trilha}: Os temas geralmente s\~ao segmentados em trilhas. Mais ainda, os ve\'iculos podem aceitar artigos com diferentes n\'iveis de maturidade, tais como resumos expandidos, artigos curtos para apresenta\c{c}\~ao de resultados preliminares de pesquisa, ou artigos completos.
  \item[$\triangleright$] \textsc{Prazos}: A submiss\~ao de artigos pode seguir em fluxo cont\'inuo ou obedecer prazos estabelecidos pelo ve\'iculo. Os autores devem observar as datas limites para submiss\~ao e os prazos de revis\~ao e publica\c{c}\~ao dados pelo ve\'iculo. N\~ao s\~ao raros os casos de ve\'iculos que levam um longo tempo para revisar e publicar o artigo, gerando atrasos indesej\'aveis na divulga\c{c}\~ao dos resultados de pesquisa.
  \item[$\triangleright$] \textsc{Normas}: Os autores devem observar os limites e restri\c{c}\~oes impostos pelas normas adotadas pelo ve\'iculo, tais como n\'umero m\'inimo e m\'aximo de p\'aginas, tamanhos de fontes, espa\c{c}amento de linhas e par\'agrafos e normaliza\c{c}\~ao bibliogr\'afica, fazendo as devidas adequa\c{c}\~oes de forma quando necess\'ario.
\end{itemize}

A segunda atividade descrita na Figura~\ref{fig:processopublicacao} \'e a defini\c{c}\~ao da estrat\'egia de submiss\~ao do artigo.
Na estrat\'egia, os autores devem considerar aspectos como a urg\^encia para publica\c{c}\~ao, o encadeamento dos cronogramas dos diversos ve\'iculos-alvo e a possibilidade de rejei\c{c}\~ao do artigo frente \`a concorr\^encia com outros artigos.
Por exemplo, os autores podem decidir submeter primeiramente para um ve\'iculo de alta reputa\c{c}\~ao e concorr\^encia, que ofere\c{c}a um tempo curto para revis\~ao e publica\c{c}\~ao, e considerar a possibilidade de submiss\~ao para outro ve\'iculo num curto intervalo de tempo em caso de rejei\c{c}\~ao, considerando a incorpora\c{c}\~ao das sugest\~oes provenientes das revis\~oes efetuadas na primeira submiss\~ao.

A terceira atividade descrita na Figura~\ref{fig:processopublicacao} consiste na efetiva\c{c}\~ao de adapta\c{c}\~oes no conte\'udo e na forma do artigo para adequa\c{c}\~ao \`as normas impostas pelo ve\'iculo. Nessa etapa, os autores devem dar aten\c{c}\~ao especial ao m\'etodo de revis\~ao adotado pelo ve\'iculo, sob pena de ter o artigo sumariamente rejeitado. Em caso de revis\~ao \`as cegas (\textit{blind review}), onde o artigo deve ser apresentado e avaliado sem a identifica\c{c}\~ao de autoria, os autores devem retirar qualquer elemento do conte\'udo do artigo que possa revelar a autoria.
Em sequ\^encia, os autores devem efetivamente submeter o artigo utilizando os meios indicados pelo ve\'iculo para tal fim.

Cabe ressaltar que o m\'etodo adotado para revis\~ao do artigo pode ser diferente para cada ve\'iculo. 
Geralmente a avalia\c{c}\~ao \'e feita por pares, ou seja, por pareceristas com um certo grau de especializa\c{c}\~ao na \'area e afinidade com o tema, e \`as cegas (\textit{blind review}).
Em caso de rejei\c{c}\~ao, alguns ve\'iculos admitem que os autores contestem a revis\~ao, apresentando argumentos que podem ser levados em considera\c{c}\~ao para que a decis\~ao de rejei\c{c}\~ao seja revertida.
Tal a\c{c}\~ao est\'a representada pela quinta atividade descrita na Figura~\ref{fig:processopublicacao}.
Independentemente da possibilidade de contesta\c{c}\~ao, no caso de rejei\c{c}\~ao do artigo pelo ve\'iculo, os autores devem rever sua estrat\'egia de submiss\~ao para decidir se realizar\~ao os ajustes sugeridos pelos revisores visando uma nova submiss\~ao para outro ve\'iculo, ou se desistir\~ao da estrat\'egia adotada e abandonar\~ao as tentativas de publica\c{c}\~ao do artigo.

Uma vez aceito para submissão, a atividade de prepara\c{c}\~ao para publica\c{c}\~ao, s\'etima e \'ultima atividade descrita na Figura~\ref{fig:processopublicacao}, deve ser executada. Nessa etapa, acordos e direitos autorais s\~ao estabelecidos entre os autores e o ve\'iculo e as \'ultimas modifica\c{c}\~oes s\~ao efetuadas, dando formato final ao artigo.

\section{Considera\c{c}\~oes Finais}
\label{sec:consideracoes}
O presente artigo reflete a experi\^encia de orienta\c{c}\~ao discente do autor e prov\^e subs\'idios para suportar a constru\c{c}\~ao e publica\c{c}\~ao de artigos cient\'ificos em computa\c{c}\~ao, especialmente os de car\'ater experimental.
Informa\c{c}\~oes complementares, tais como dicas, links, materiais e sugest\~oes de ferramentas e m\'etodos para escrita podem ser encontradas na p\'agina \href{http://www.icei.pucminas.br/professores/wladmir/artcom/}{Escrevendo Artigos Cient\'ificos em Computa\c{c}\~ao}.

\nocite{Zobel:2004@Springer}
%\nocite{Olson:2014@EA}
\nocite{Hall:2013@Wiley}
\nocite{Day:1998@Oryx}

%\nocite{EC}

\bibliographystyle{plain}
\bibliography{references}

\end{multicols}
\vspace{30pt}
\begin{minipage}{.3\textwidth}
\includegraphics[scale=1.845]{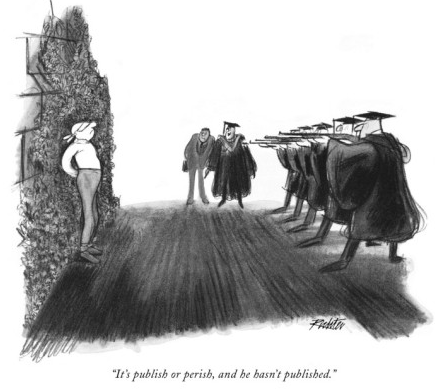}
\captionof{figure}{Charge do \textit{The New Yorker}, 1966.}
\label{fig:publishperish}         
\end{minipage}
\begin{minipage}{.7\textwidth}
\centering
\includegraphics[scale=2]{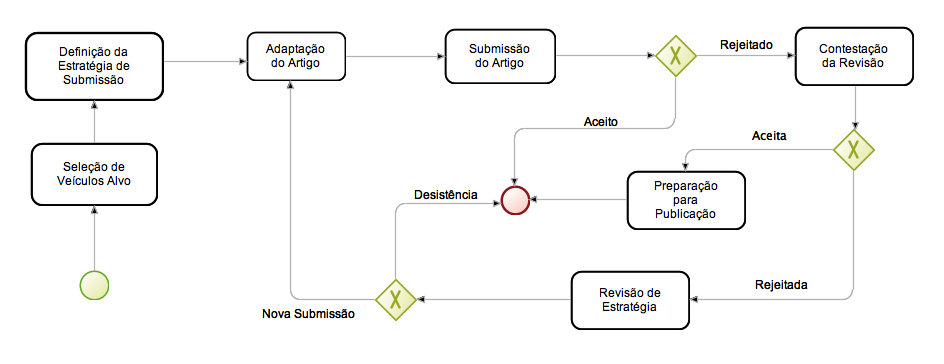}
\captionof{figure}{Processo convencional de publica\c{c}\~ao de artigo cient\'ifico em computa\c{c}\~ao.}
\label{fig:processopublicacao}            
\end{minipage}%

% \begin{figure*}[b]
% \centering
% \includegraphics[scale=2.25]{processopublicacao.png}
% \caption{Processo de publica\c{c}\~ao de artigo cient\'ifico em computa\c{c}\~ao.}
% \label{fig:processopublicacao}
% \end{figure*}

\end{document}